# The nonequilibrium statistical operator with finite duration of the present time moment

V. V. Ryazanov

Institute for Nuclear Research, Kiev, pr.Nauki, 47, Ukraine;
e-mail: vryazan@kinr.kiev.ua

The method of the nonequilibrium statistical operator accounts for the history of a system, influence of its past history to its present state. It is suggested to take into account the finite duration of the present time moment ("continuous present"), which according to I.Prigogine's results, is equal to the average Lyapunov time.

In the previous paper [1] we interpreted the logarithm of the nonequilibrium statistical operator (NSO) $\ln\rho_1(t)$ as an averaging of the logarithm of the quasi-equilibrium statistical operator $\ln\rho_q$ over the distribution of the lifetime of a system $p_q^{(0)}(u)$:

$$\ln\rho(t)=\ln\rho_1(t)=\int_0^\infty p_q^{(0)}(u)\ln\rho_q(t-u,-u)du = \int_{-\infty}^t p_q^{(0)}(t-t_0)\ln\rho_q(t_0,t_0-t)dt_0 =$$

$$= \ln\rho^R(t) = \ln\rho_q(t,0) - \int_0^\infty (\int p^{(0)}{}_q(u)du)\frac{\partial}{\partial u}\ln\rho_q(t-u,-u)du , \qquad (1)$$

where for the case of NSO in the form by Zubarev [2-3] $\rho_{zub}$ $p_q^{(0)}(u)=\varepsilon_1\exp\{-\varepsilon_1 u\}$, $u=t-t_0$. This exponential distribution results in the fact that $\rho_1=\rho_{zub}$ and yields a linear relaxation source in the Liouville equation in the form $J=-\varepsilon_1(\ln\rho(t)-\ln\rho_q(t,0))$, $\varepsilon_1=1/\langle\Gamma\rangle$, where $\langle\Gamma\rangle=\langle t-t_0\rangle$ is the average lifetime of a system (up to the present moment t). This source can be also included in the Liouville operator, giving its modified form suggested by Prigogine [4] (see also [5]) for the dynamic conditions of dissipativity and introducting irreversibility at the microscopic level.

The operation of averaging over lifetime corresponds to the introduction of the operator of internal time in the approach of [4]. Averaging with another (not exponential) distribution for $p^{(0)}{}_q(u)$ gives another form of the source and dissipative part of Liouville operator [6]. The exponential distribution is the only one which possesses the Markovian property [7]: regardless to the age, the remaining lifetime does not depend on the past and has the same distribution as the lifetime itself. Thus $\varepsilon_1^{-1}=\langle t-t_0\rangle$, and $\langle t_f-t_0\rangle=2/\varepsilon_1$, where $t_0$ and $t_f$ are random moments of birth and destruction of the system, t is the present time moment, Fig.1.

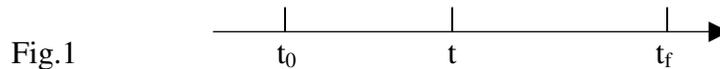

Fig.1    $t_0$    t    $t_f$

The NSO in the form (1) considers the time $t-t_0$, time lived by the system up to a present moment t, that is the past of the system. The state of the system at the time moment t is influenced by the contribution of states from its last history, that is from the interval of receiving the information t-t₀. Therefore in (1) $\rho^R$ denotes the retarded solution. It is possible to write the advanced solution as well

$$\ln \rho^A(t) = \int_0^\infty p^{(f)}{}_q(u) \ln \rho_q(t+u,u) du. \qquad (2)$$

Full lifetime of a system equals $t_f-t_0$, and the interval $t_f-t$ seems also to play some part in the evolution of the system and in its current state. In [2] it is marked that the superposition of advanced (considering the future time moments) and retarded solutions is observed for nonequilibrium systems where the fluxes reflected from borders are essential.

Similar situation takes place in the problems of radiation where the standing waves are obtained by the superposition of retarded and advanced potentials. Generally the value $p_q^{(0)}(u)$ need not to be in the form of the exponential distribution; the exponential form itself is an approximation and a limiting case [8]. Possibilities of other choices of $p_q^{(0)}(u)$ and their impact on the kinetics of system are considered in [6].

Other distributions $p_q^{(0)}(u)$ except exponential do not imply the Markovian property and a complete description of a system with finite size needs considering the interval $t_f-t$ (and not only the part $t-t_0$) in the systems with lifetime $t_f-t_0$; the latter is an essential physical characteristic of the system affecting its behaviour.

In [4] the contributions from the past and future time moments are considered within the notion of the internal time. Only the past moments and the moments from the "nearest future" are shown to contribute to the present state of a system. Therefore the notion of "continuous present" is introduced in [4], that is the transitive layer from the past to the future having width $\tau_L$ where $\tau_L^{-1}$ is Lyapunov average exponent.

In order to take into account the influence of the continuous present in NSO, it is necessary to perform the time averaging over the duration of the continuous present and to use the superposition of retarded (1) and advanced (2) solutions of the kind

$$\ln \rho_2(t) = \int_0^\infty p^{(f)}{}_q(y_f)(\int_0^\infty p^{(0)}{}_q(u) \ln \rho_q(t+y_f-u, y_f-u)du)dy_f =$$

$$= \int_0^\infty p^{(0)}{}_q(u)(\int_0^\infty p^{(f)}{}_q(y_f) \ln \rho_q(t+y_f-u, y_f-u)dy_f)du = <\ln\rho^R(t)>_f = <\ln\rho^A(t)>_0 = \qquad (3)$$

$$= \ln\rho_q(t,0) - \int_0^\infty (\int p^{(0)}{}_q(u)du)(\frac{d}{du}\ln\rho_q(t-u,-u))du - \int_0^\infty (\int p^{(f)}{}_q(y_f)dy_f)(\frac{d}{dy_f}\ln\rho_q(t+y_f, y_f))dy_f +$$

$$+ \int_0^\infty (\int p^{(f)}{}_q(y_f)dy_f) \int_0^\infty (\int p^{(0)}{}_q(u)du) \frac{d}{dy_f}\frac{d}{du} \ln\rho_q(t+y_f-u, y_f-u)du)dy_f; \quad y_f = t_f - t.$$

The "suppression" of the future states and the account for the contribution from the "nearest" future only is reached by a corresponding choice of the average value $<\Gamma_{fut}> = \int_0^\infty y_f p_q^{(f)}(y_f) dy_f \sim \tau_L$, distinct from both $<t-t_0>$ and $<t_f-t>$. We consider different "times of living": the duration of the period of reception of the information from the past with an average value $<t-t_0>$, corresponding to the past history of a system; and the average time $<\Gamma_{fut}>$ - the duration of the period of reception of the information from the future or from the continuous future.

A general form of the source J in the Liouville equation

$$\frac{\partial \rho}{\partial t} + iL\rho(t) = J; \quad \rho = \rho_2 \qquad (4)$$

for the distribution in the form (3) looks like

$$J = p^{(0)}{}_q(0)(\int_0^\infty p^{(f)}{}_q(y_f) \ln \boldsymbol{\rho}_q(y_f + t, y_f) dy_f) +$$

$$+ \int_0^\infty \frac{\partial p^{(0)}{}_q(u)}{\partial u} (\int_0^\infty p^{(f)}{}_q(y_f) \ln \boldsymbol{\rho}_q(t + y_f - u, y_f - u) dy_f) du -$$

$$- \int_0^\infty p^{(0)}{}_q(u) p^{(f)}{}_q(0) \ln \boldsymbol{\rho}_q(t - u, -u) du - \qquad (5)$$

$$- \int_0^\infty p^{(0)}{}_q(u) (\int_0^\infty \frac{\partial p^{(f)}{}_q(y_f)}{\partial y_f} \ln \boldsymbol{\rho}_q(t + y_f - u, y_f - u) dy_f) du -$$

$$- \int_0^\infty p^{(0)}{}_q(u) (\int_0^\infty p^{(f)}{}_q(y_f) \frac{\partial}{\partial t} \ln \boldsymbol{\rho}_q(t + y_f - u, y_f - u) dy_f) du.$$

The NSO in a form (3) can also be cast in the form:

$$\ln \rho_2(t) = \ln \rho^{(R)}(t) - \Delta; \qquad (6)$$

$$\Delta = \int_0^\infty (p^{(f)}{}_q(y_f) dy_f) \int_0^\infty p^{(0)}(u) \frac{d}{dy_f} \ln \boldsymbol{\rho}_q(t + y_f - u, y_f - u) du dy_f =$$

$$= \int_0^\infty [p^{(0)}(0) \ln \boldsymbol{\rho}_q(t + y_f, y_f) du + \int_0^\infty \frac{\partial p^{(0)}(u)}{\partial u} \ln \boldsymbol{\rho}_q(t + y_f - u, y_f - u) du](\int p^{(f)}(y_f) dy_f) dy_f$$

For

$$p^{(0)}(u) = \varepsilon_0 \exp\{-\varepsilon_0 u\}; \quad p^{(f)}(u) = \varepsilon_f \exp\{-\varepsilon_f u\}: \qquad (7)$$

$$\ln \rho_2(t) = \ln \rho^{(R)}(t) + \Delta_1; \qquad (8)$$

$$\Delta_1 = \frac{\varepsilon_0 / \varepsilon_f}{1 + \varepsilon_0 / \varepsilon_f} [\ln \rho^{(A)}(t) - \ln \rho^{(R)}(t)]; \quad \frac{\varepsilon_0 / \varepsilon_f}{1 + \varepsilon_0 / \varepsilon_f} = \frac{\tau_L}{\tau_L + <t - t_0>}.$$

The terms $\Delta_1$ are different from zero if: a) $\varepsilon_0 \neq 0$, $\varepsilon_f \neq 0$; б) $\varepsilon_0 \to 0$, $\varepsilon_f \to 0$, (Boltzmann-Grad limit [4]). The term $\Delta_1 \to 0$ if both $\varepsilon_f \to \infty$, $\tau_L \to 0$ and at bigger times, when $\varepsilon_0 \to 0$.

The expression of the source for $\rho_2(t)$ in a form (7), (8) is

$$J = -\varepsilon_0 [\ln \rho_2 - \ln \rho^{(A)}(t)] = \varepsilon_f [\ln \rho_2 - \ln \rho^{(R)}(t)].$$

Function $\rho_2(t)$ relax to $\rho^{(A)}(t)$ with average time $<t - t_0>$ or to $\rho^{(R)}$ with time $\tau_L$.

The expression (3) for a case (7) can also be written in the form of (1), but the integration should be performed between $-\infty$ and $\infty$, and the lifetime distribution function $p_q^{(0)}(u)$ is replaced with the function of the kind

$$\ln \rho_2(t) = \int_{-\infty}^{\infty} p^{(0f)}{}_q(u) \ln \rho_q(t-u,-u)du \ ; \ p_q^{(0f)}(u)=(1+z)^{-1}(\chi_{u>0}\varepsilon_0\exp\{-\varepsilon_0 u\}+ z\chi_{u<0} \quad (9)$$

$\varepsilon_f\exp\{\varepsilon_f u\})$; $\chi_{u>0}=1$, $u>0$; $\chi_{u>0}=0$, $u<0$; $\chi_{u<0}=1$, $u<0$, $\chi_{u<0}=0$, $u>0$; $z=\varepsilon_0/\varepsilon_f=\tau_L/<t-t_0>$.

The expressions (9) can also be interpreted as including the "history" $t-t_0$ with average $\varepsilon_0=<t-t_0>^{-1}$ and the "prehistory" $t_0-t$, the time before the system birth with average $\varepsilon_f=\tau_L^{-1}$, in the description of a system.

The amendments to $\ln \rho_1 = \ln \rho^{(R)} = \ln \rho_{zub}$ are

$$\ln \rho_2 = \ln \rho_1 - \frac{z}{1+z}(\int_0^{\infty} e^{-\varepsilon_0 u} + \int_{-\infty}^{0} e^{\varepsilon_f u})\frac{\partial}{\partial u}\ln \rho_q(t-u,-u)du . \quad (10)$$

The fracture $z = \varepsilon_0/\varepsilon_f = \tau_L/<t-t_0> >> 1$ for "young" systems when the values $<t-t_0>$ are small. In particular, this is the case of the for nucleation, drop formation, forming structures. Changing the value of the amendment to $\ln \rho_q(t,0)$ in (9) - (10) leads to changing the form of the transport equations and other nonequilibrium characteristics (kinetic coefficients, etc.), as it is shown in [6]. It is possible to write these changes in a general fashion using the results of [2-3] and [9]. For the case of the non-isothermic nucleation the estimations of the values of these amendments in the method of NSO was performed in [10].